\documentclass[10pt]{article}

\usepackage[T1]{fontenc}
\usepackage[utf8]{inputenc}
\usepackage[english]{babel}
\usepackage{amsmath,latexsym,amsfonts,amsthm,bm,mathtools}
\usepackage{graphicx}
\usepackage[top=1.5in, bottom=1.5in, left=1.25in, right=1.25in]{geometry}
\usepackage{booktabs,url,tikz}

\title{{ssMousetrack}: Analysing computerized tracking data via Bayesian state-space models in {R}}
\author{Antonio Calcagn\`{i}\footnote{Corresponding author: antonio.calcagni@unipd.it},
	Massimiliano Pastore, and 
	Gianmarco Alto\`{e}\\ \\
 University of Padova 
}

\date{}

\DeclareMathAlphabet{\mathcall}{OMS}{zplm}{m}{n}
\newcommand{\timeN}[1]{{(#1)}}
\newcommand{\fct}[1]{\texttt{#1()}}
\newcommand{\pkg}[1]{\textbf{#1}}

\begin{document}

\maketitle

\begin{abstract}
Recent technological advances have provided new settings to enhance individual-based data collection and computerized-tracking data have became common in many behavioral and social research. By adopting instantaneous tracking devices such as computer-mouse, wii, and joysticks, such data provide new insights for analysing the dynamic unfolding of response process. \pkg{ssMousetrack} is a {R} package for modeling and analysing computerized-tracking data by means of a Bayesian state-space approach. The package provides a set of functions to prepare data, fit the model, and assess results via simple diagnostic checks. This paper describes the package and illustrates how it can be used to model and analyse computerized-tracking data. A case study is also included to show the use of the package in empirical case studies.\\\vspace{0.15cm}

\end{abstract}

\section{Introduction}

Recent technological advances allow the collection of detailed data on ratings, attitudes, and choices during behavioral tasks. Unlike standard surveys and questionnaires, these tools provide a rich source of data as they adopt tracking devices that collect subject-based information about the dynamics involved during the data collection task \cite{freeman+al:2010,schulte-mecklenbeck+al:2019}. Examples of such devices include eye-tracking, body movement-tracking, computer mouse-tracking, and electrodermal activity. Among these, computer mouse-tracking has become an important and widely used tool in behavioral sciences, as it provides a valid and cost-effective way to measure the usually unknown processes underlying human ratings and decisions \cite{freeman:2017}. Mouse-tracking data are often collected by means of standard computer mouse, wii instruments, and joystick devices and consist of collections of real-time trajectories recorded during the behavioral task. In a typical mouse-tracking task, individuals are presented with a computer-based interface showing the stimulus at the bottom of the screen (e.g., the image of a ``dolphin'') and two labels on the left and right top corners (e.g., the labels ``mammal'' vs. ``fish''). They are asked to decide which of the two labels is appropriate given the task instruction and stimulus (e.g., to decide whether dolphin is mammal or fish). In the meanwhile, the x-y trajectories of the computer device are instantaneously recorded. The real-time trajectories offer an effective way to study the decision process underlying the hand movement behavior by revealing, for instance, the presence of some levels of decisional uncertainty. The applications of mouse-tracking tools spread across many research area, including cognitive sciences \cite{coco+duran:2016}, neuroscience \cite{stolier+freeman:2017}, neurology \cite{ruitenberg+al:2016}, and forensic studies \cite{monaro+al:2017}. 

Several tools for running mouse-tracking analyses are available in open-source specialized software like MouseTracker \cite{freeman+al:2010}, EMOT \cite{calcagni+al:2017}, and MouseTrap \cite{kieslich+henninger:2017}. In the {R} environment, only the packages \pkg{mousetrack} \cite{Coco+al:2015} and \pkg{mousetrap} \cite{Kieslich+al:2018} are devoted to mouse-tracking data. More generally, there are other packages developed to handle with tracking data such as \pkg{trajectories} \cite{pebesma+al:2015}, \pkg{trackeR} \cite{frick+al:2017}, \pkg{adehabtatLT} \cite{calenge2006package}, and \pkg{move} \cite{kranstauber+al:2017}. Similarly, there are many packages developed for state-space models like, for instance, \pkg{KFAS} \cite{helske:2017} and \pkg{bssm} \cite{helske+vihola:2018}. 

In this paper we present \pkg{ssMousetrack}, a novel {R} package to analyse computerized tracking data as they emerge from typical mouse-tracking data recording. The package implements a non-linear state space model to handle with the dynamics of mouse-tracking data. The model is estimated using approximated Kalman filter coupled with MCMC algorithms via the \pkg{rstan} package \cite{rstan:2018}. The package includes functions for data pre-processing, data generation, and model assessment. It also provides functionalities to set-up designs for mouse-tracking data recording. Despite other {R} packages are available in this context, \pkg{ssMousetrack} differs in some aspects. For instance, \pkg{mousetrack} and \pkg{mousetrap} focus on descriptive evaluation of mouse-tracking data static measures (e.g., minima, maxima, flips, curvature). By contrast, our package (i) implements a dynamic evaluation of the trajectory data without resorting the use of summary measures and (ii) evaluates the observed data variability in terms of latent dynamics and external covariates (e.g., experimental variables), which are usually of relevance in mouse-tracking data analysis. With respect to \pkg{KFAS} and \pkg{bssm}, our package offers a more focused implementation for mouse-tracking data. Moreover, although complete and useful in many cases, these packages implement a general class of state-space representations which may not be widely applicable to computerized tracking data. Finally, \pkg{ssMousetrack} differs also from \pkg{trajectories}, \pkg{trackeR}, \pkg{adehabtatLT}, and \pkg{move} as they focus on animal tracking and related problems, such as estimation of habitat choices. This makes them not directly suitable for analysing the various aspects of mouse-tracking studies. In this respect, the advantages of \pkg{ssMousetrack} are as follows: (i) it is easy to use as it requires typing a single function to run the entire procedure, (ii) it takes advantages of \pkg{rstan} package to estimate model parameters via MCMC, (iii) it allows modeling and analysing computerized tracking data as they are usually recorded in typical mouse-tracking tasks, (iv) it provides a user-friendly workflow for all processing steps which can be easily understood by non-expert users, (v) it offers users a way to simulate mouse-tracking designs and data as well. In addition, \pkg{ssMousetrack} can be combined with other {R} packages, including \pkg{shinystan} \cite{shinystan:2018} and \pkg{ggmcmc} \cite{ggmcmc:2016} to produce further statistical and graphical representation of the output. The package is available from the Comprehensive {R} Archive Network at \url{https://CRAN.R-project.org/package=ssMousetrack}. 

This paper is organized as follows. Section 2 gives a statistical overview of the model implemented in the package, the methods of estimation and inference, and the assessment of the model. Section 3 describes the package's structure and its utilities. Section 4 illustrates the functioning of the package by means of an illustrative case study. Finally, Section 5 concludes the manuscript with a discussion and future directions. 

\section{Model}\label{sec:model}

Computerized mouse-tracking data typically consist of arrays $(\mathbf x, \mathbf y)_{ij} \in \mathbb{R}^{N_{ij}} \times \mathbb{R}^{N_{ij}}$ containing the streaming of x-y Cartesian coordinates of the computer-mouse pointer, for $i=1,\ldots,I$ subjects, $j=1,\ldots,J$ stimuli, and $n=1,\ldots,N_{ij}$ time steps. To simplify data analysis, raw trajectories are usually pre-processed according to the following steps \cite{hehman+al:2014,calcagni+al:2017}. First, the trajectories $(\mathbf x, \mathbf y)_{ij}$ are normalized on a common sampling scale such that $N$ is the same over subjects and stimuli. Next, the arrays $(\mathbf x, \mathbf y)_{ij} \in \mathbb{R}^{N} \times \mathbb{R}^{N}$ are rotated and translated into the quadrant $[-1,1]\times [1,1]$ with $(x_0,y_0)_{ij} = (0,0)$ and $(x_N,y_N)_{ij} = (1,1)$ by convention. Finally, normalized data are projected onto a (lower) 1-dimension space via \textit{atan2} function. In this way, the final ordered data $\mathbf{y}_{ij} = \big(y^\timeN{1},\ldots,y^\timeN{n},\ldots,y^\timeN{N}\big) \in (0,\pi]^N$ lie on the arc defined by the union of two disjoint sets, i.e. the set $\{y \in (0,\pi]: y \geq \frac{\pi}{2}\}$ which represents the right-side section of the screen (usually called \textit{target}, T) and the set $\{y \in (0,\pi]: y < \frac{\pi}{2}\}$ which instead represents the left-side section (usually called \textit{distractor}, D). The final data are arranged as an $\mathbf{Y}_{N\times JI}$ column-wise stacked matrix. 

The state-space model implemented in \pkg{ssMousetrack} is as follows:
\begin{align}
\label{eq1a} \mathbf y^\timeN{n}_{JI\times 1} \sim \text{vonMises}\Big( \bm \mu^\timeN{n}_{JI\times 1}, \bm \kappa^\timeN{n}_{JI\times 1} \Big)\\[0.2cm]
\label{eq1b} \bm \mu^\timeN{n}_{JI\times 1} = \mathcal{G}\Big( \mathbf{x}_{I\times 1}^\timeN{n}, \bm{\beta}_{J\times 1}\Big)\\[0.2cm]
\label{eq1c} \mathbf{x}^\timeN{n}_{I\times 1} \sim \text{Normal}\Big(\mathbf{x}^\timeN{n-1}_{I\times 1},\sigma_x\mathbf{I}_{I\times I}\Big)\\[0.2cm]
\label{eq1d} \bm\beta_{J\times 1} = \mathbf{Z}_{J\times K} \cdot \bm\gamma_{K\times 1} \\[0.2cm]
\label{eq1e} \bm\kappa_{JI\times 1}^\timeN{n} = \exp^\dagger\Big(\lambda\mathbf d_{JI\times 1}^\timeN{n}\Big)
\end{align}
where Equation \eqref{eq1a} is a von Mises \textit{measurement equation} with $\bm \mu^\timeN{n}_{JI\times 1} \in (0,\pi]^{IJ}$ being the mean for the $n$-th time step and $\bm \kappa^\timeN{n}_{JI\times 1} \in \mathbb{R}^{IJ}$ the concentrations around the $n$-th mean vector, Equation \eqref{eq1b} represents the locations on the arc defined in $(0,\pi]$ from which the data vector $\mathbf y^\timeN{n}$ is sampled from and it behaves according to the a real function $\mathcal{G}: \mathbb{R} \to (0,\pi]$, which maps reals into radians. In the current version of \pkg{ssMousetrack}, $\mathcal{G}$ can be defined as:
\begin{itemize}
	\item[(i)] $\pi$-scaled logistic function: 
	\begin{equation*}
	\mathcal{G} = \text{vec}\Bigg(\pi^{-1}\Big[1 + \exp\Big(\bm\beta_{J\times 1}\mathbf{1}_{1\times I} - \mathbf{1}_{J\times 1}\mathbf x_{1\times I}^\timeN{n} \Big)\Big) \Big] \Bigg)
	\end{equation*}
	\noindent with $\bm\beta_{J\times 1} \in \mathbb{R}^{J}$ representing the contribution of the stimuli on $\mathbf y^\timeN{n}$.
	\item[(i)] $\pi$-scaled Gompertz function:
	\begin{equation*}
	\mathcal{G} = \text{vec}\Bigg(\pi\Big[\exp\Big( -\bm\beta_{J\times 1}\mathbf{1}_{1\times I} \exp\Big( \mathbf{1}_{J\times 1}\mathbf x_{1\times I}^\timeN{n} \Big)\Big) \Big] \Bigg)
	\end{equation*}
	\noindent where $\bm\beta_{J\times 1} \in \mathbb{R}_+^{J}$ has the same meaning as before.
\end{itemize}
Although they represent two cases of the general family of $S$-shaped functions, logistic and Gompertz models differ in some respects. For instance, unlike the logistic model, the Gompertz function is not symmetric around its inflection point, with the consequence that its growth rises rapidly to its maximum rate occurring before the fixed inflection point \cite{mcneish+dumas:2017}. Moreover, the parameters of the Gompertz function are always positive, a constrain which is often required by applications where the covariates of the model cannot take negative values (e.g., reaction times). These two implementations allow users to choose the type of $\mathcal G$ function on the basis of the experimental designs they have used in their studies.

In Equation \eqref{eq1c} is a Normal \textit{states equation} which represents a lag-1 autoregressive process with time-fixed variance parameter $\sigma_x$. In the current version of \pkg{ssMousetrack}, the covariance matrix of the latent processes is set to an identity matrix $\mathbf{I}$ without loss of generality ($\sigma_x = 1$). Equation \eqref{eq1d} is the linear term modeling the contribution of the experimental design (e.g., two-by-two design) and variables involved (e.g., categorical variables, continuous covariates). Note that $\mathbf{Z}$ is a design (dummy) matrix of main and high-order effects defined by adopting the dummy coding (e.g., treatment contrasts, sum contrasts) whereas $\bm\gamma$ is the associated vector of parameters for the columns of $\mathbf{Z}$, with $\gamma_1$ being the usual baseline term for the contrasts. Finally, Equation \eqref{eq1e} defines the concentrations around the $n$-th location by using the transformed data:
\begin{equation*}
\mathbf d^\timeN{n} = 
\begin{cases}
\big|\mathbf y^\timeN{n} - \frac{3\pi}{4}\big|, &~\text{if}~ \mathbf y^\timeN{n} < \frac{\pi}{2}\\[0.2cm]
\big|\mathbf y^\timeN{n} - \frac{\pi}{4}\big|, &~\text{if}~ \mathbf y^\timeN{n} \geq \frac{\pi}{2}\\	 	 
\end{cases}
\end{equation*}
with $\exp^\dagger : (0,\pi] \to [\text{lb},\text{ub}] \subset \mathbb{R}_+$ being the exponential function scaled in the natural range of the concentration parameter (e.g., $lb=10$, $ub=200$). In the current implementation of the package, the parameter $\lambda$ is fixed to unity.

The interpretation of Equations \eqref{eq1b}-\eqref{eq1d} is as follows. The $n$-th mean vector $\bm\mu^\timeN{n}$ is expressed as function of the stimuli-related component $\bm\beta$ and subject-based component $\mathbf x^\timeN{n}$, which are integrated together to form the conditional sampling $\mathbf{y}^\timeN{n}|\bm\beta,\mathbf{x}^\timeN{n}$ through the function $\mathcal G$. As a result, Equation \eqref{eq1c} can be interpreted as the \textit{individual latent dynamics} that are unaffected by the experimental stimuli whereas Equation \eqref{eq1d} represents the \textit{experimental effect} regardless to individual dynamics. More generally, Equation \eqref{eq1c} conveys information about the hand movement process underlying the tracking device and as such it can be used to analyse how much individuals differ in executing the task. By contrast, Equation \eqref{eq1d} collects information on how a certain experimental manipulation has an effect or not on the movement responses. Interestingly, when normalized into $[0,1]$, $\bm\mu^\timeN{n}$ can be interpreted as the probability of the $i$-th individual at the $j$-th stimulus to navigate close the distractor cue in the left-side section of the arc. Finally, Equation \eqref{eq1e} follows from the fact that hand movements underlying computerized tracking data tend to be smooth over the experimental task, with small changes being more likely close to left (distractor) or right (target) endpoints \cite{brockwell+al:2004}.

\subsection{Estimation and inference}

The state-space model in Equations \eqref{eq1a}-\eqref{eq1e} requires estimating the array of latent trajectories $\mathbf{X} \in \mathbb{R}^{I\times N}$ together with the array of parameters $\bm\gamma_{K\times 1}$, with $\gamma_1 \in \mathbb{R}$ and $\bm\gamma_{(K-1)\times 1} \in \mathbb R^{K-1}$ (logistic case) or $\bm\gamma_{(K-1)\times 1} \in [-\gamma_1,\infty)^{K-1}$ (Gompertz case). The array of unknown quantities $\bm\Theta = \{\mathbf{X},\bm\gamma\}$ can be estimated in various way, by adopting both a frequentist and Bayesian perspectives \cite{shumway+stoffer:2006}. In the \pkg{ssMousetrack} package, the parameters are recovered in Bayesian way by means of a marginal MCMC algorithm through which $\mathbf{X}$ and $\bm\gamma$ are alternately recovered \cite{andrieu+al:2010,sarkka:2013}. The reason is twofold: (i) MCMC algorithms, as those implemented in \pkg{rstan} package, provide a more efficient and complete
solution for sampling from the probability distribution of the parameters. (ii) The Bayesian approach offers an elegant solution for data analysis and inference \cite{gelman+al:2014} by means of which the model is adequately assessed by the analysis of (marginal) posterior distributions of the parameters \cite{kruschke:2014}. 

More in details, the posterior density $f(\bm\Theta|\mathbf Y)$ after factorization of the joint density $f(\bm\gamma,\mathbf{X},\mathbf Y)$, is as follows \cite{andrieu+al:2010}:
\begin{equation}\label{eq2}
f(\bm\Theta|\mathbf Y) \propto f(\bm\gamma) f(\bm\gamma|\mathbf{Y}) f(\mathbf X|\mathbf{Y})
\end{equation}
where $f(\bm\gamma|\mathbf{Y})$ is the (marginal) likelihood function, $f(\mathbf X|\mathbf{Y})$ is the filtering density, whereas $f(\bm\gamma)$ is the prior ascribed on the model parameters. In the current version of \pkg{ssMousetrack}, $f(\mathbf X|\mathbf{Y})$ is approximated via Kalman filtering/smoothing, with $f(\bm\gamma|\mathbf{Y})$ being computed as a byproduct of the Kalman theory (see Appendix \ref{app1}). 

\subsection{Model assessment}\label{sec:evalmodel}

In the Bayesian context of data analysis, \pkg{ssMousetrack} provides a simulation-based procedure to evaluate the adequacy of the model to reproduce the observed data $\mathbf Y$ \cite{gelman+al:2014}. More technically, given the posteriors of parameters and latent states $f(\bm\Theta|\mathbf Y)$, $M$ new (simulated) datasets $\mathbf{Y}^*_1,\ldots,\mathbf Y^*_M$ are generated according to the estimated model structure and, for each new dataset, two discrepancy measures are considered \cite{kiers1997techniques}:
\begin{align}
\label{eq3a} \text{PA}_\text{overall} = 1 - \Big(||\mathbf Y_m - \mathbf Y||^2 \big/ ||\mathbf Y||^2\Big)\\
\label{eq3b} \text{PA}_\text{sbj} = 1 - \Big(||\mathbf Y^{(i)}_m - \mathbf Y^{(i)}||^2 \big/ ||\mathbf Y^{(i)}||^2\Big) \\
\quad i=1,\ldots,I\nonumber
\end{align}
which measure the total amount of data reconstruction based on the overall $JI \times N$ observed array $\mathbf Y$ (Equation \ref{eq3a}) and the amount of data reconstruction based on the $J \times N$ observed matrix $\mathbf Y^{(i)}$ for each subject $i = 1,\ldots,I$ (Equation \ref{eq3b}). Both the indices are in the range 0-100\%, with 100\% indicating optimal fit. Note that the measure $\text{PA}_\text{sbj}$ allows for evaluating the adequacy of the model to reconstruct the individual-based set of data. In addition, the \textit{dynamic time warp distance} (dtw), as implemented in \pkg{dtw} package, is also computed between $\mathbf Y^{(i)}_m$ and $\mathbf Y^{(i)}$. Unlike the $\text{PA}_\text{sbj}$ index, the dtw distance measures the similarity among time series by considering their different dynamics \cite{giorgino2009computing}.

\section{The ssMousetrack package}

The \pkg{ssMousetrack} is distributed via the Comprehensive R Archive Network (CRAN). It is based on \pkg{rstan} \cite{rstan:2018}, the {R} interface to the probabilistic programming language Stan \cite{carpenter+al:2017}. The current version of the package allows for (i) simulating data according to a given experimental design, (ii) analysing mouse-tracking data via state-space modeling, and (iii) evaluating the adequacy of model results. The package consists of five main function (\fct{generate\_data}, \fct{run\_ssm}, \fct{check\_prior}, \fct{prepare\_data}, \fct{evaluate\_ssm}), two datasets (\texttt{language}, \texttt{congruency}), and three sub-functions (\fct{compute\_D}, \fct{generate\_Z}, \fct{generate\_design}). The main functions \fct{generate\_data} and \fct{run\_ssm} are wrappers to previously-compiled Stan codes which implement the model described in Section \ref{sec:model}. Table \ref{tab1} provides an overview of the functions and datasets provided in the \pkg{ssMousetrack} package whereas a description of the usage of the functions is reported in the next subsections. 

\begin{table}[!t]
	\centering
	\begin{tabular}{llp{9.5cm}}
		function & type & description \\\toprule
		\fct{generate\_data} & main & simulate data according to a user-defined experimental design. \\		
		\fct{run\_ssm} & main & run state-space model on a given mouse-tracking dataset. \\
		\fct{check\_prior} & main & allows users to define a list of priors for $f(\bm\gamma)$ prior running \fct{run\_ssm}. \\
		\fct{prepare\_data} & main & pre-process raw tracking data prior running \fct{run\_ssm}.   \\
		\fct{evaluate\_ssm} & main & run model evaluation given an output of \fct{run\_ssm}. The function can plot results if requested by users. \\				
		\fct{compute\_D} & internal & compute the matrix of distances $\mathbf D$ given the observed data $\mathbf Y$ (see Equation \ref{eq1e}). \\
		\fct{generate\_Z} & internal & generate the Boolean trial-by-variable (design) matrix $\mathbf Z$ (see Equation \ref{eq1d}). \\
		\fct{generate\_design} & internal & allows users to specify an experimental design in terms individuals, trials, variables, and design matrix $\mathbf Z$. \\
		\texttt{congruency} & dataset & subset of data from \cite{coco+duran:2016}.\\
		\texttt{language} & dataset & subset of data from \cite{barca+pezzullo:2012}.\\\bottomrule
	\end{tabular}
	\caption{\label{tab1} Overview of the contents of \pkg{ssMousetrack}}
\end{table}

\subsection{Generate artificial data}

To simulate artificial data we use the function \fct{generate\_data}, which requires as input the experimental template for the data generation process. More generally, the function works by first sampling the parameters $\bm\gamma$ from the prior density $f(\bm\gamma)$ and then generates the latent states $\mathbf X$ from Equation \eqref{eq1c}, computes the matrix $\bm\mu$ from Equation \eqref{eq1b} and the matrix $\mathbf D$, drawns the matrix of data $\mathbf Y$ by simply applying Equation \eqref{eq1a}. For instance, an experiment with one categorical independent variable and two levels, each with three trials, can be generated via the following syntax:

\begin{verbatim}
	prior_list <- list("normal(-0.25,0.5)","normal(2.7,1)")
	datagen1_ssm <- generate_data(N = 61, M = 100,I = 2,J = 6,
	+                                K = 2,Z.formula = "~Z1", priors = prior_list)
\end{verbatim}

where \texttt{M = 100} is the number of data to be generated, \texttt{N = 61} is the number of time step for the mouse-tracking trajectories, \texttt{K = 2} means that we have just one variable with two levels, \texttt{J = 6} indicates the total number of trials such that \texttt{J/K} is the number of trials for each level of the variable, \texttt{I = 2} is the number of subject, \texttt{Z.formula} indicates the formula for the contrast matrix $\mathbf Z$ with standard {R} syntax. Note that selective priors are specified for each level of the experimental design using the Stan syntax (see the help of \fct{check\_prior} for further details).

The output is a list containing three sublists, as follows:
\begin{itemize}
	\item \texttt{params}, which contains the model parameters generated for the $M$ datasets:
	\begin{verbatim}
		## List of 4
		##  $ sigmax: num [1:2] 1 1
		##  $ lambda: num [1:12] 1 1 1 1 1 ...
		##  $ gamma : num [1:75, 1:2] 0.228 -0.378 ...
		##  $ beta  : num [1:75, 1:12] 0.228 -0.378 ...
	\end{verbatim}
	
	\item \texttt{data}, which contains the matrices of latent states $\mathbf X$ and trajectories $\mathbf Y$, together with $\bm\mu$, $\mathbf D$, and $\mathbf Z$: 
	
	\begin{verbatim}
		## List of 5
		##  $ Y : num [1:75, 1:61, 1:12] 1.54 1.53 ...
		##  $ X : num [1:75, 1:61, 1:2] 1e-04 1e-04 1e-04 1e-04 1e-04 ...
		##  $ MU: num [1:75, 1:61, 1:12] 1.57 1.57 ...
		##  $ D : num [1:75, 1:61, 1:12] 0.785 0.785 ...
		##  $ Z : num [1:12, 1:2] 1 1 1 1 1 ...
		\end{verbatim}
		
		\item \texttt{design}, which contains the experimental design used as template to generate the data:
		
		\begin{verbatim}
		##   sbj trial  Z1
		## 1   1     1 100
		## 2   1     2 100
		## 3   1     3 100
		\end{verbatim}
		
		\end{itemize}
		
		Similarly, artificial datasets can be generated using more complex designs. For instance, a bivariate design with two variables is produced by typing:
		
		\begin{verbatim}
		datagen2_ssm <- generate_data(I = 2,J = 8,K = c(2,4),Z.formula = "~Z1*Z2",
		+                         Z.type=c("symmetric","random"))
		\end{verbatim}
		where \texttt{K = c(2,4)} codifies two variables each with two and four levels, \texttt{Z.formula = "~Z1*Z2"} indicates that the variables interact whereas \texttt{Z.type=c("symmetric","random")} indicates that trials must be assigned to the first variable using the symmetric method and to the second variable using the random method (see the help of \fct{generate\_Z} for further details). 
		
\begin{figure}[h!]
		\centering
		\includegraphics[width=14.8cm]{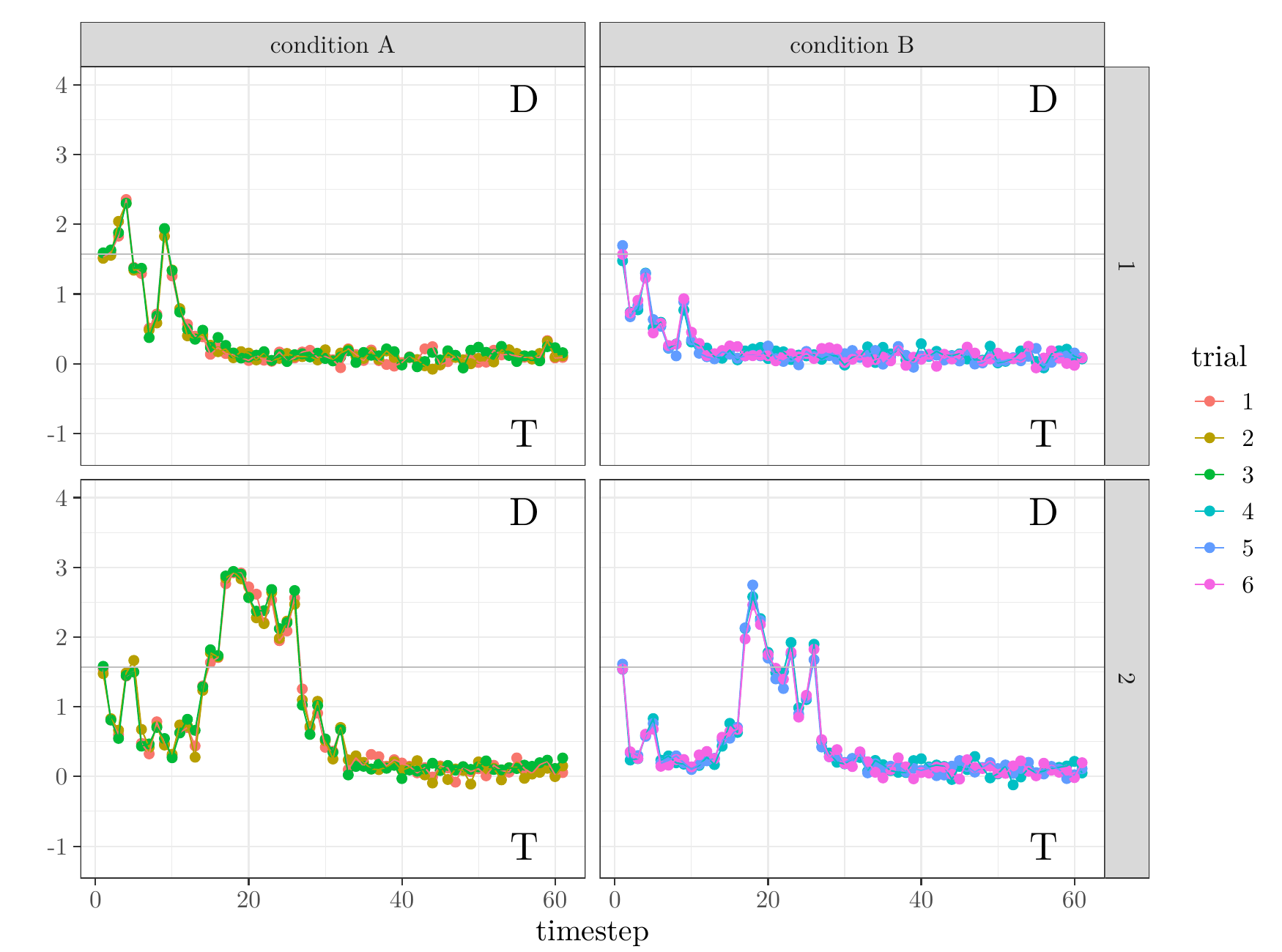} 	
		\caption{Simulated mouse-tracking trajectories $\mathbf Y$ plotted over time step $n=1,\ldots,N$. The data refer to a univariate design with $I=2$ (rows of the plot), $K=2$ (columns of the plot) and $J = 6$ (colors of the plot). Note that each of the $K$ levels has $J/K$ trials, T and D indicate Target and Distractor sections of the movement space.}
		\label{fig:fig1}
		\end{figure}

		Figure \ref{fig:fig1} shows a sample of mouse-tracking data $\mathbf Y$ generated in the univariate design case with \texttt{I = 2}, \texttt{K = 2} and \texttt{J = 6}. We report the univariate case only for the sake of simplicity but the same graphical representations can be done for the more complex designs as well.
		
		\subsection{Run state-space analysis}
		
		State-space analysis can be run on both simulated and real data. In the first case, after the data-generation process, the state-space model implemented in the \pkg{ssMousetrack} package can be fit using \fct{run\_ssm}. For instance, the syntax:
		
		\begin{verbatim}
		datagen2_ssm <- generate_data(I = 2,J = 8,K = c(2,4),Z.formula = "~Z1*Z2",
		+                         Z.type=c("symmetric","random"))
		iid <- 2
		datagen2_fit <- run_ssm(N = datagen2_ssm$N,I = datagen2_ssm$I,
		+                          J = datagen2_ssm$J,Y = datagen2_ssm$data$Y[iid,,],
		+                          D = datagen2_ssm$data$D[iid,,],Z = datagen2_ssm$data$Z,
		+                          niter = 5000,nwarmup = 2000,nchains = 2)
	\end{verbatim}
	runs the state-space analysis on the \texttt{iid = 2} artificial data \texttt{datagen2\_ssm}. Note that \texttt{niter} indicates the number of total samples to be drawn, \texttt{nwarmup} the number of warmup/burnin iterations per chain, and \texttt{nchains} the number of chains to be executed in parallel. The function \fct{run\_ssm} allows for parallel computing via the \pkg{parallel} package when \texttt{nchains} > 1. In this case, since \texttt{ncores="AUTO"} (default), the function will run two parallel chains using two cores. 
	
	Unlike for the case of artificial data, the analysis of real datasets requires preparing raw data in a proper format via \fct{prepare\_data}, the function that implements the steps described in Section \ref{sec:model}. Generally, raw datasets need to be organized using the long-format, with information being organized as nested. The dataset \texttt{language} is an example of a typical data structure required by \fct{prepare\_data}:
	
	\begin{verbatim}
		data("language")
		str(language,vec.len=2)
		## 'data.frame':	6060 obs. of  6 variables:
		##  $ sbj      : int  1 1 1 1 1 ...
		##  $ trial    : int  1 1 1 1 1 ...
		##  $ condition: Factor w/ 4 levels "HF","LF","PW",..: 1 1 1 1 1 ...
		##  $ timestep : int  1 2 3 4 5 ...
		##  $ x        : num  0 -0.0098 -0.0098 -0.0098 -0.0098 ...
		##  $ y        : num  0 0.0025 0.0025 0.0025 0.0025 ...
	\end{verbatim}
	
	where \texttt{condition} is the categorical variable involved in the study. The pre-processing of raw data is performed by the call:
	
	\begin{verbatim}
		language_proc <- prepare_data(X = language,N = 61,Z.formula = "~condition")
	\end{verbatim}
	
	where the output \texttt{language\_proc} is a data frame containing the pre-processed dataset together with the column-wise stacked matrix $\mathbf Y$ of angles, the contrast matrix $\mathbf Z$, and the matrix of distances $\mathbf D$. 
	
	Once raw data have been pre-processed, the state-space analysis is performed as for the case of artificial data: 
	
	\begin{verbatim}
		language_fit <- run_ssm(N = language_proc$N,I = language_proc$I,
		+                          J = language_proc$J,Y = language_proc$Y,
		+                          D = language_proc$D,Z = language_proc$Z,
		+                          niter = 5000,nwarmup = 2000,nchains = 2)
	\end{verbatim}
	
	The function returns as output a list composed of three sublists, as follows:
	\begin{itemize}
		\item \texttt{params}, which contains the posterior samples for the free parameters $\bm\gamma$ and $\bm\beta$:
		
		\begin{verbatim}
			## List of 6
			##  $ sigmax    : num 1
			##  $ lambda    : num 1
			##  $ kappa_bnds: num [1:2] 5 300
			##  $ gamma     :'data.frame':	4000 obs. of  4 variables:
			##  $ beta      : num [1:4000, 1:60] -0.26 -0.146 ...
			##  $           :function (z, ...)
		\end{verbatim}
		
		\item \texttt{data}, which contains the posterior samples for the latent states $\mathbf X$ and the moving means $\bm\mu$:
		
		\begin{verbatim}
			## List of 6
			##  $ Y       : num [1:101, 1:60] 1.56 1.7 ...
			##  $ X       : num [1:4000, 1:101, 1:5] 1e-04 1e-04 1e-04 1e-04 1e-04 ...
			##  $ MU      : num [1:4000, 1:101, 1:60] 1.76 1.68 ...
			##  $ D       : num [1:101, 1:60] 0.592 0.474 ...
			##  $ Z       : num [1:60, 1:4] 1 1 1 1 1 ...
			##  $ X_smooth: num [1:4000, 1:101, 1:5] -0.0878 -0.0635 ...
		\end{verbatim}
		
		\item \texttt{stan\_table}, containing the typical Stan output (i.e., point estimates, credibility intervals, and Gelman-Rubin index) for the \fct{sampling} method as implemented in the \pkg{rstan} package:
		
		\begin{verbatim}
			##           mean se_mean   sd  2.5%   25%   50%  75% 97.5% n_eff Rhat
			## gamma[1] -0.05       0 0.19 -0.43 -0.18 -0.05 0.08  0.33  3047    1
			## gamma[2] -0.02       0 0.06 -0.13 -0.06 -0.02 0.02  0.09  2764    1
			## gamma[3]  0.16       0 0.06  0.04  0.12  0.16 0.20  0.28  2782    1
		\end{verbatim}
		
	\end{itemize}
	
	Note that users can also export the \texttt{stanfit} object with all the Stan results by specifying \texttt{stan\_object=TRUE} in \fct{run\_ssm}. 
	
	The function \fct{run\_ssm} allows for different priors specification. In particular, users can specify different priors for the model parameters $\bm\gamma$ as follows:
	
	\begin{verbatim}
		priors_list <- list("lognormal(1,0.5)","normal(1,2)","chi_square(2)","normal(0,1)")
		language_fit <- run_ssm(..., priors = priors_list)
	\end{verbatim}
	which means that $\gamma_1 \sim \text{lognormal}(1,0.5)$, $\gamma_2 \sim \text{normal}(1,2)$, $\gamma_3 \sim \text{chi\_square}(2)$, $\gamma_4 \sim \text{normal}(0,1)$. The list of probability distributions accepted by \fct{run\_ssm} is described in the help of the function \fct{check\_prior}. Specification of priors for single parameters is also allowed, by using \texttt{NULL} attributes:
	
	\begin{verbatim}
		priors_list <- list(NULL,"normal(1,2)","chi_square(2)",NULL)
		language_fit <- run_ssm(..., priors = priors_list)
	\end{verbatim}
	where predefined priors are used for parameters $\gamma_1$ and $\gamma_4$. Further examples about \fct{run\_ssm} are illustrated in the manual of the package.
	
	\subsection{Evaluate the model results}
	
	The methods described in Section \ref{sec:evalmodel} for the model evaluation are implemented by the function \fct{evaluate\_ssm}, which requires as input the output of \fct{run\_ssm}. For instance, considering the fitted object \texttt{language\_fit}, the model evaluation can simply be run via the command:
	
	\begin{verbatim}
		language_eval <- evaluate_ssm(ssmfit = language_fit, M = 1000, plotx = FALSE)
	\end{verbatim}
	where \texttt{M = 1000} is the number of replications to compute the indices. The function returns as ouput a list containing the mean values of the indices $\text{PA}_\text{overall}$, $\text{PA}_\text{sbj}$, and \textit{dtw}, as well as the distributions obtained over the \texttt{M} replications. Note that, users can also ask for a graphical representation of the indices by setting \texttt{plotx = TRUE}.

\section{An Illustrative example}\label{sec:examples}

In this section we provide a full example about the way \pkg{ssMousetrack} can be used for state-space analysis of real computerized tracking data. Note that the application reported here has an illustrative purpose only. To this end, we will make use of the dataset \texttt{language}, a subset of data originally presented in \cite{barca+pezzullo:2012}. In this typical computerized tracking task, participants saw a printed stimulus on the screen (e.g., the word \textit{water}) and were requested to perform a dichotomous choice task where stimuli need to be classified as word or non-word. The experimental variable \texttt{condition} was a categorical variables with four levels (\texttt{HF}: High-frequency word; \texttt{LF}: Low-frequency word; \texttt{PW}: Pseudo-word; \texttt{NW}: Non-word). Participants had to classify each stimulus as word vs. non-word by using a computer-mouse tracking device. The dataset contains $I=5$ participants, $J=12$ trials, one categorical variable with $K=4$ levels, each with $J/K=3$ trials. From the data-analysis viewpoint, we evaluat the extent to which the parameters of the state-space model $\bm\gamma$ reflect eventual differences associated with the levels of \texttt{condition}.

\begin{figure}[h!]
	\centering
	\includegraphics[width=12cm]{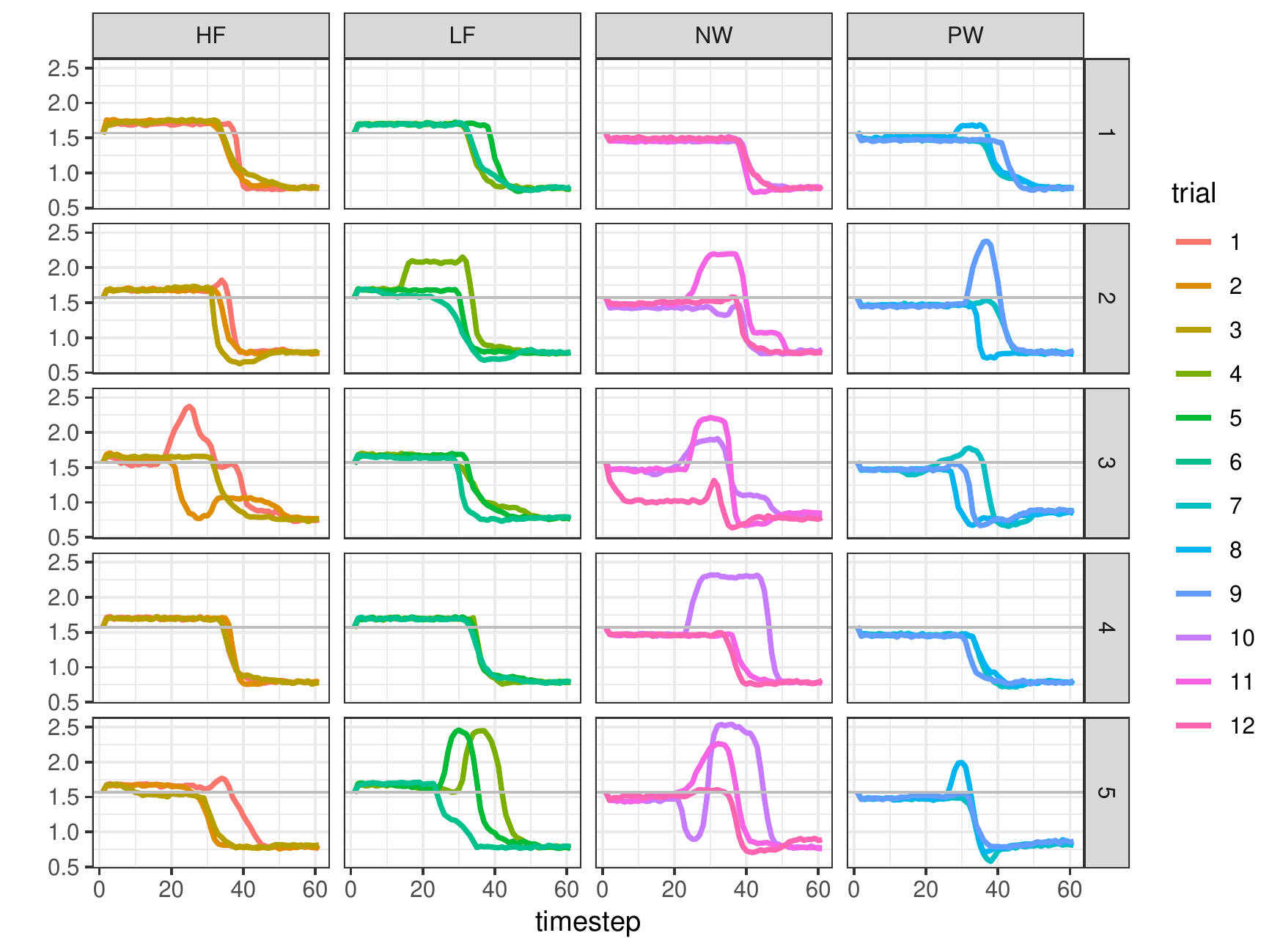}		
	\caption{\texttt{language} dataset: Mouse-tracking trajectories $\mathbf Y$ plotted over the time step $n=1,\ldots,N$. Note that the categorical levels are represented column-wise, subjects are represented row-wise, whereas distractor (D) and target (T) sections are represented above and below the solid gray line.}
	\label{fig:fig4} 
\end{figure}

The raw computerized tracking trajectories in the dataset consist of Cartesian coordinates with $N=101$ ($i=1,\ldots,I$; $j=1,\ldots,J$). The dataset is partially pre-processed as raw trajectories have the same length ($N=101$). However, we need to run \fct{prepare\_data} in order to rotate/translate the raw data into the quadrant $[-1,1]\times [1,1]$ and compute the \textit{atan2} projections. The pre-processing step is called by the command:

\begin{verbatim}
	data("language")
	language_proc <- prepare_data(X = language, N = 101, Z.formula = "~condition")
\end{verbatim}
Figure \ref{fig:fig4} shows the trajectories $\mathbf Y$ associated with the task for all participants and trials.

\begin{figure}[t!]
	\centering		
	\includegraphics[width=13cm]{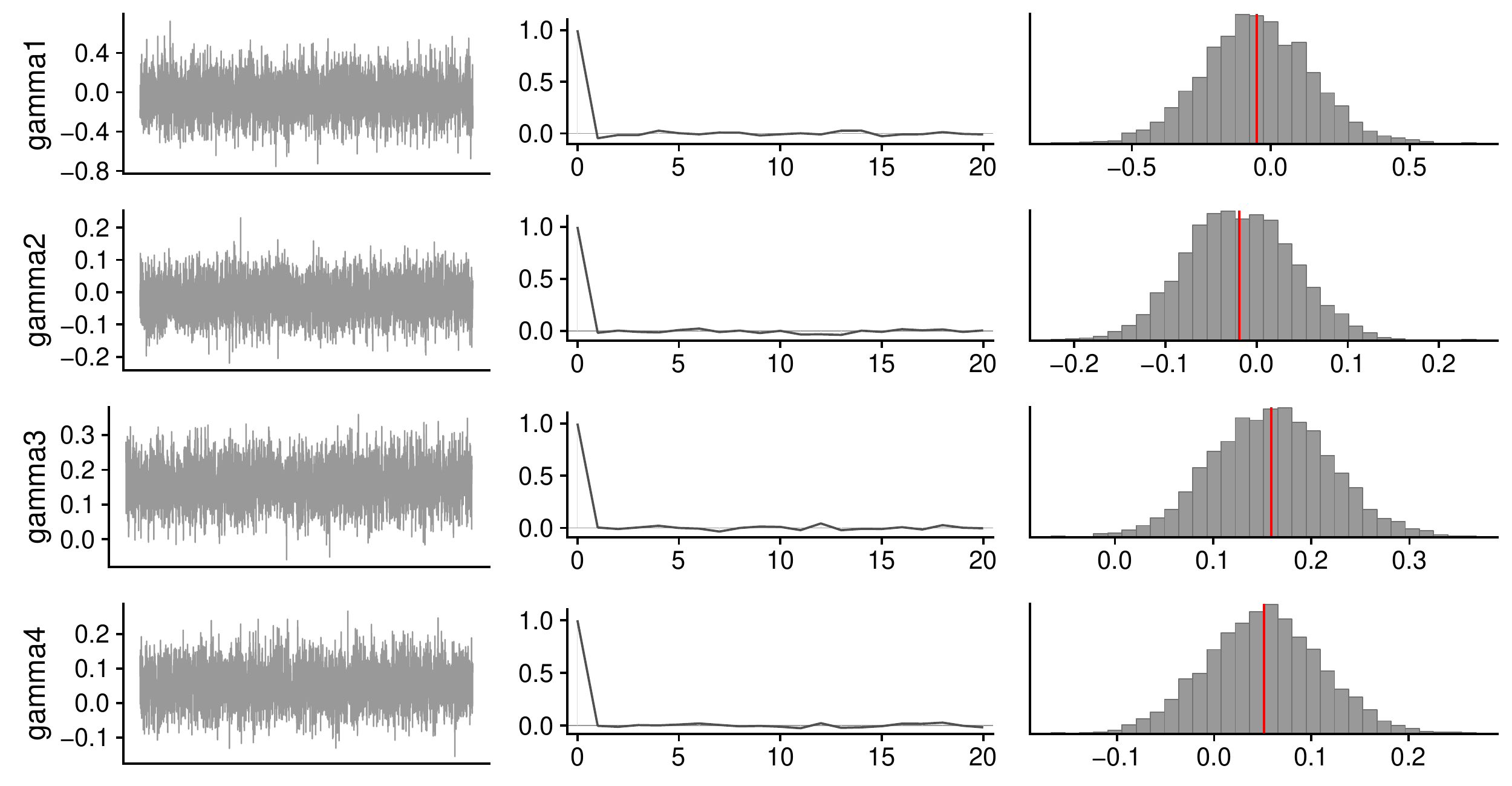} 
	\caption{\label{fig:fig5} Illustrative example: MCMC traces, autocorrelation plots, and marginal posterior distributions for the model parameters. Note that all the Gelman-Rubin indices (Rhat) of the parameters are 1.0.}
\end{figure}

\begin{figure}[h!]
	{\centering \includegraphics[width=14cm]{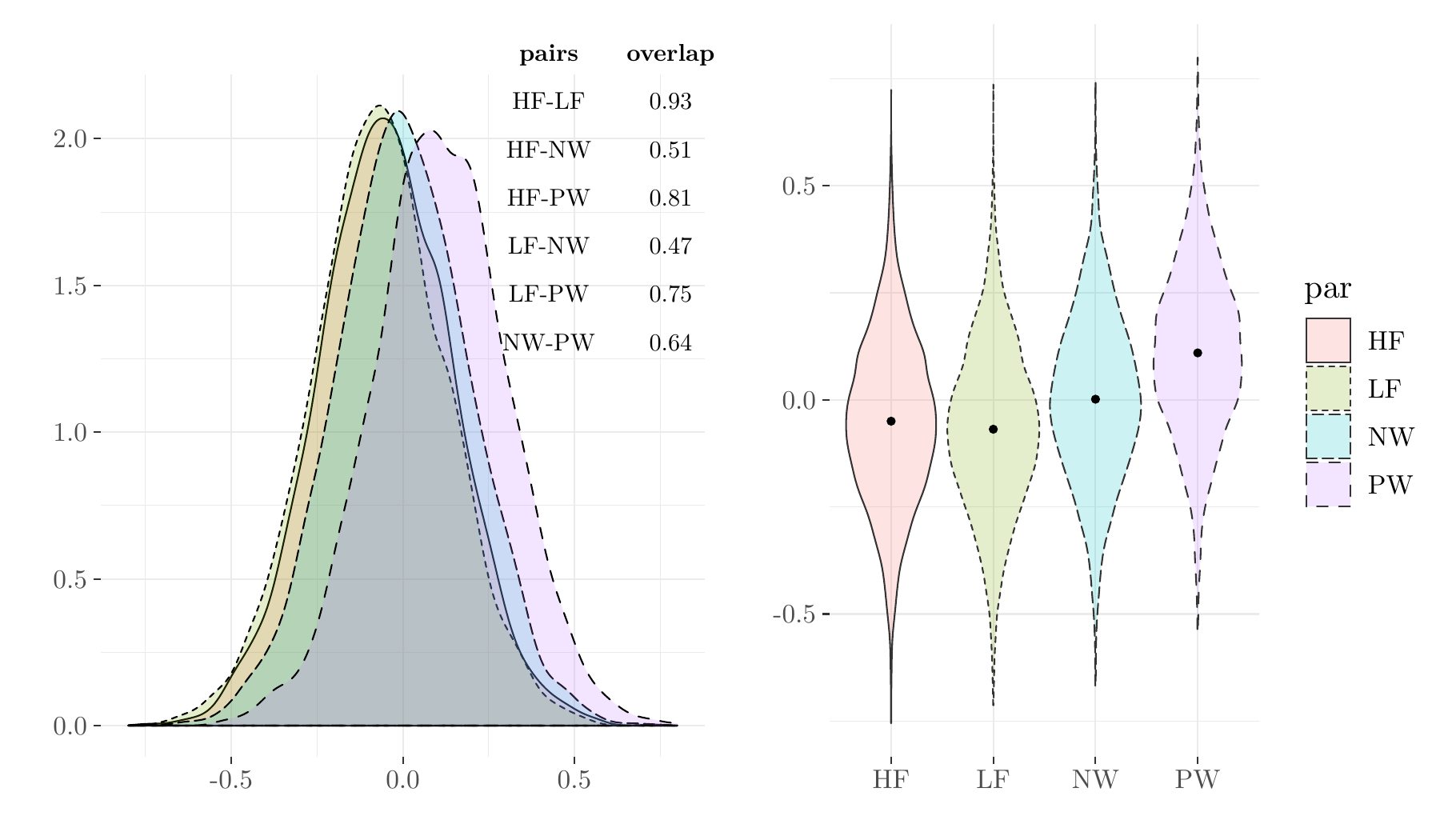} 	
	}
	\caption{Illustrative example: Marginal posterior densities and violin plots for the levels of \texttt{condition} (recoded via dummy code). Note that black dots represent posterior means of the parameters $\bm\gamma$ whereas overlaps have been computed via the package \pkg{overlapping} \cite{pastore:2018}.}
	\label{fig:fig7}
\end{figure}

Next, the state-space model is fit to the pre-processed data by the following call:

\begin{verbatim}
	priors_list <- list("normal(0,1)","normal(1,1)","normal(-2,1)","normal(2,1)")
	language_fit <- run_ssm(N = language_proc$N,I = language_proc$I,
	+                          J = language_proc$J,Y = language_proc$Y,
	+                          D = language_proc$D,Z = language_proc$Z,
	+                          niter = 6000,nwarmup = 2000,nchains=4,
	+                          priors = priors_list,
	+                          gfunction = "logistic")
\end{verbatim}
where, in this case, the prior for $\bm\gamma$ have been choosen to codify a priori expectations about the effect of the variable \texttt{condition} \cite{barca+pezzullo:2012}. Figure \ref{fig:fig5} shows some MCMC graphical diagnostics for the model parameters $\bm\gamma$ computed using \pkg{bayesplot} \cite{bayesplot:2018} whereas Table \ref{tab:tab2} reports the posterior quantities for the model parameters. In the Bayesian context of data-analysis, we evaluate the effects of the variable \texttt{condition} by computing the degree of overlapping among marginal posterior densities for each level of the experimental variable (i.e., the more the overlapping, the weaker the evidence supporting the experimental manipulation). Figure \ref{fig:fig7} shows the results graphically. Overall, the variable \texttt{condition} showed no strong effect, as the densities of the levels are overlapped. In particular, stimuli in HF, LF, and NW conditions showed no activation of the distractor section of the tracking space as $\hat{\gamma}_\text{HF}$, $\hat{\gamma}_\text{NW}$, and $\hat{\gamma}_\text{LF}$ approach zero. By contrast, stimuli in PW condition showed a small effect on activating the target section ($\hat{\gamma}_\text{HF} > 0$), possibly due to the fact that PW stimuli require less cognitive workload \cite{barca+pezzullo:2012}.

\begin{table}[ht]
	\centering
	\begin{tabular}{rrrrrrrr}
		\hline
		& mean & sd & 25\% & 50\% & 75\% & n\_eff & Rhat \\ 
		\toprule
		gamma1 & -0.05 & 0.19 & -0.18 & -0.05 & 0.08 & 3047.00 & 1.00 \\ 
		gamma2 & -0.02 & 0.06 & -0.06 & -0.02 & 0.02 & 2764.00 & 1.00 \\ 
		gamma3 & 0.16 & 0.06 & 0.12 & 0.16 & 0.20 & 2782.00 & 1.00 \\ 
		gamma4 & 0.05 & 0.06 & 0.01 & 0.05 & 0.09 & 2680.00 & 1.00 \\ 
		\bottomrule
	\end{tabular}
	\caption{Illustrative example: Posterior quantities and Gelman-Rubin indices (Rhat) for the model parameters.} 
	\label{tab:tab2}
\end{table}

\begin{figure}[h!]
	{\centering \includegraphics[width=15cm]{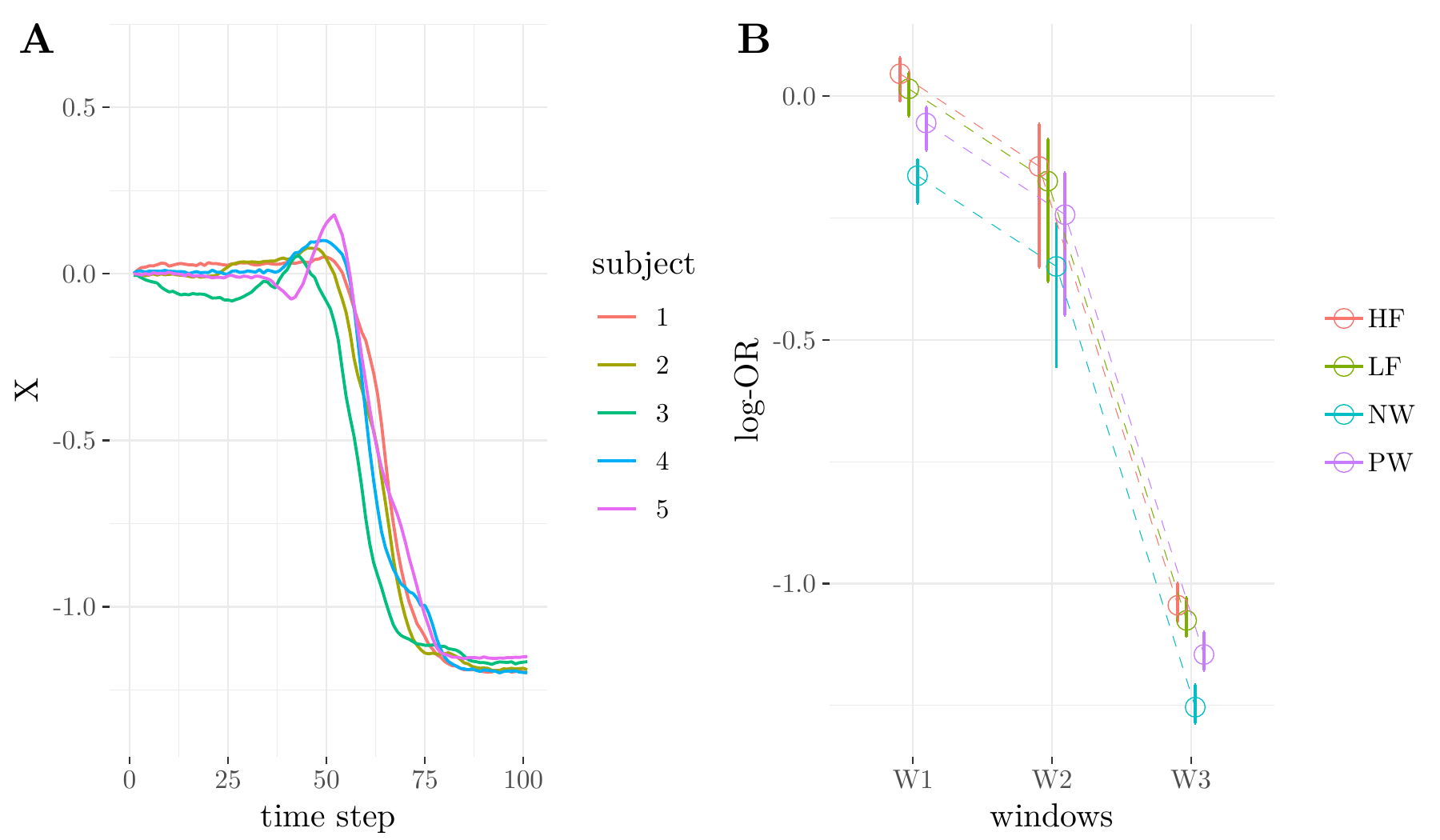}}
	\caption{Illustrative example: (A) Estimated latent dynamics $\mathbf{\hat{X}}$ for $I=5$ subjects plotted over the scale $N=0\%,\ldots,100\%$. (B) log-Odd ratio for the evidence analysis.}
	\label{fig:fig6} 
\end{figure}

Figure \ref{fig:fig6}-A reports the filtered latent states $\mathbf{\hat{X}}$ for the subjects in the dataset. To further investigate how individual dynamics differ over the levels of \texttt{condition}, we can make use of $\mathbf{\hat{X}}$ and ask whether HF, LF, NW, and PW stimuli differ in terms of evidence of mouse-tracking competition. The idea is that the higher the evidence, the larger the difficulty in categorizing stimuli as word (target) or non-word (distractor). 

To do this, we follow the findings of \cite{barca+pezzullo:2012} and divide the entire respose process $1,\ldots,N$ into three disjoint windows $W_1=10-35\%$, $W_2=45-65\%$, and $W_3=70-85\%$. Usually it is expected that a higher competition would be observed in $W_1$ and $W_2$ rather than $W_3$. More formally, let $\mathbf{\hat{x}}^\timeN{i}_{M\times 1} = (\hat{x}_1^\timeN{i},\ldots,\hat{x}_M^\timeN{i})$ be the sequence of filtered states for the $i$-th subject and the generic time window $W$, with $M$ being equals to the cardinality of $W$. Next, the probability to select non-word (distractor) responses are computed by normalizing the $\mathcal G$ function into the domain $[0,1]$, as follows:
\begin{equation}
\mathbf P^\timeN{i}_{M\times K} = \Big[1+\exp\Big\{\mathbf{\hat{x}}^\timeN{i}_{M\times 1}\mathbf{1}_{1\times K} - \mathbf{1}_{M\times 1}\bm{\hat{\gamma}}_{1\times K} \Big\}\Big]^{-1}
\end{equation}
where $\bm{\hat{\gamma}}$ is the array of posterior means of the model parameters. Note that in this example we use the logistic function because we set \texttt{gfunction="logistic"} in \fct{run\_ssm}. Finally, the evidence measures can be defined in terms of log-odd ratio using the probability matrix $\mathbf P^\timeN{i}$: 
\begin{equation}
\mathbf{r}^\timeN{i}_{K\times 1} = \log\Big( {\mathbf{{p}}^\timeN{i}_{K\times 1}}\Big/{1-\mathbf{{p}}^\timeN{i}_{K\times 1}} \Big)
\end{equation}
where $\mathbf p_{K\times 1} = \frac{1}{M} \Big( \mathbf 1_{1\times M} \mathbf P^\timeN{i}_{M\times K} \Big)^T$ is the profile probability for HF, LF, NW, and PW. The interpretation of $\mathbf{r}^\timeN{i}$ is as follows. For $\mathbf{r}^\timeN{i} > 0$ there is a higher competition in categorizing the stimulus as word (target) vs. non-word (distractor). By contrast, for $\mathbf{r}^\timeN{i} < 0$ there is a lower competition in the response process, as stimuli are easily categorized as word (target). Finally, the case $\mathbf{r}^\timeN{i} = 0$ indicates that there is no difference in terms of evidence between word (target) and non-word (distractor) responses. Figure \ref{fig:fig6}-B shows the results for the four levels of \texttt{condition}. As expected, the competition in the third phase of the response process $W_3$ is low, as the probability to select the target is higher. The same applies to $W_2$. On the contrary, in the first stage of the process $W_1$ the competition is higher although the evidence ratio for all the levels of \texttt{condition} approximate zero. Interestingly, the second phase $W_2$ shows a higher whithin-subject variability of competition, which probably indicates that subjects differ in the categorization process just in the middle phase of the response process.

Finally, we assess the adequacy of the model with regards to the observed data by means of \fct{evaluate\_ssm}, as follows:

\begin{verbatim}
	language_fit_eval <- evaluate_ssm(ssmfit = language_fit,M = 500,plotx = FALSE)
\end{verbatim}
where \texttt{language\_fit} is the fitted object returned by \fct{run\_ssm} whereas \texttt{M = 500} is the number of replications used to compute the three fit indices. The output of the function consists of a list containing means and distributions of the fit indices:

\begin{verbatim}
	## List of 2
	##  $ dist   :List of 3
	##   ..$ PA_ov : num [1:500] 0.944 0.938 ...
	##   ..$ PA_sbj: num [1:500, 1:5] 0.991 0.991 ...
	##   ..$ DTW   : num [1:500, 1:60] 0.0945 0.1105 ...
	##  $ indices:List of 3
	##   ..$ PA_ov : num 0.936
	##   ..$ PA_sbj: num 0.99
	##   ..$ DTW   : num 0.119
\end{verbatim}

\begin{figure}[h!]
	{\centering \includegraphics[width=14cm]{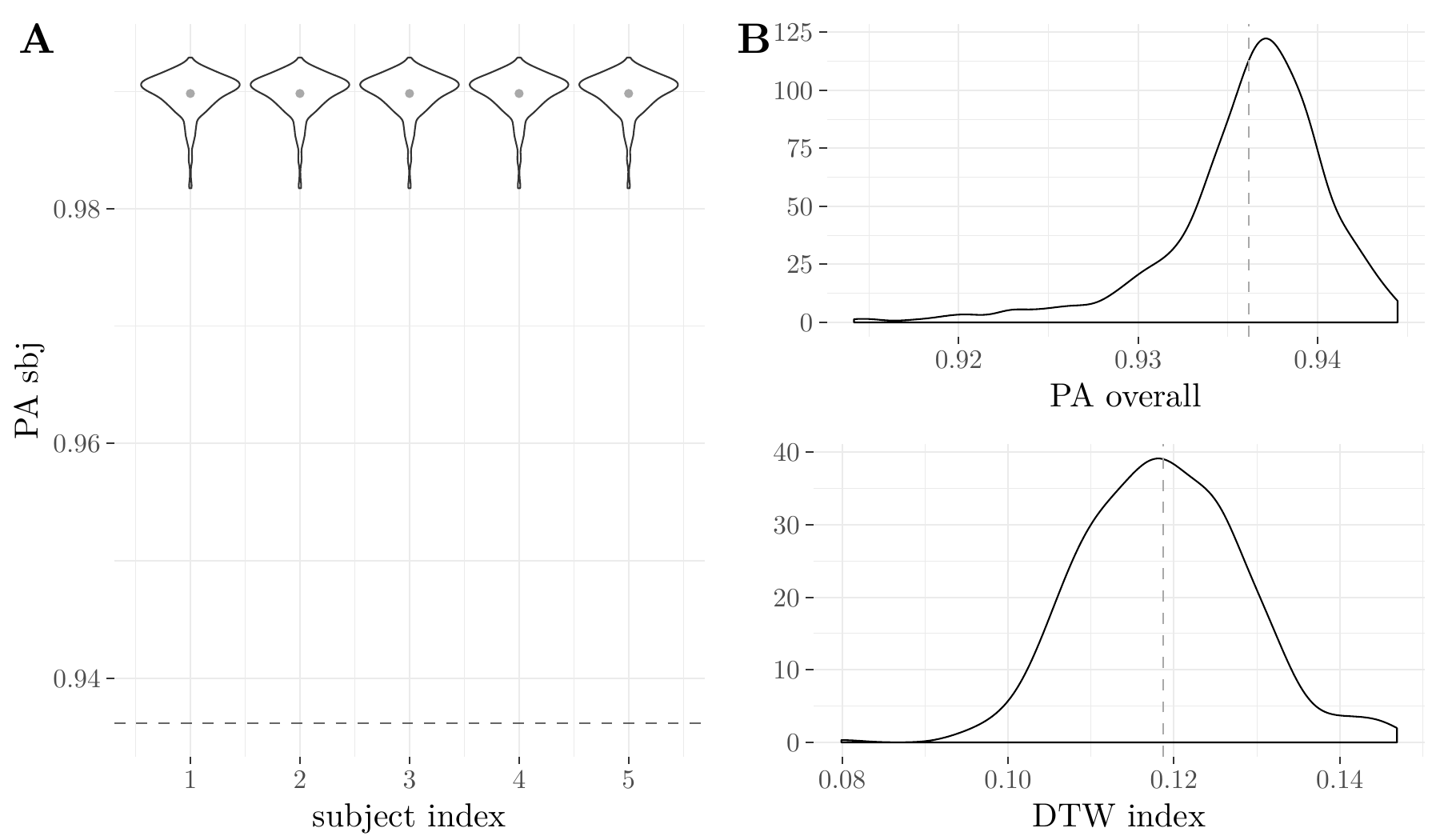}}		
	\caption{Illustrative example: Distributions of $\text{PA}_\text{ov}$, $\text{PA}_\text{sbj}$, and \textit{dtw} fit indices. Note that dotted lines represent the average $\text{PA}_\text{ov}$.}
	\label{fig:fig8} 
\end{figure}

Overall, in this example the fitted model is adequate to reproduce the observed trajectory data as supported by high values of the indices $\text{PA}_\text{ov}$, $\text{PA}_\text{sbj}$, and \textit{dtw}. Figure \ref{fig:fig8} shows the results graphically.

\section{Conclusions}

In this paper we introduced the {R} package \pkg{ssMousetrack} that analyses computerized-tracking data using Bayesian state-space modeling. The package provides a set of functions to facilitate the preparation and analysis of tracking data and offers a simple way to assess model fit. The package can be of particular interest to researchers needing tools to analyse computerized-tracking experiments using a complete statistical modeling environment instead of descriptive statistics only. In addition, the package \pkg{ssMousetrack} allows for individual-based analysis of trajectories where latent dynamics are used to obtain richer information which can pave the way to further analyses (e.g., profile analysis). The current version of the package can be extended in several ways. For instance, the inclusion of other state-space representations beyond the simple Gaussian AR(1) model can be a further generalization of the package. Still, model parameters like $\sigma_x$ and $\lambda$ can be free to allow for multi-group analysis. Similarly, more comprehensive model diagnostics could also be considered in future releases of the package. 

Finally, we believe our package may be a useful tool supporting researchers and practitioners who want to make analysis of computerized-tracking experiments using a statistical modeling environment. This will surely help them to improve the interpretability of data analysis as well as the reliability of conclusions they can draw from their studies.

\clearpage
\begin{appendix}
	\section{Appendix} \label{app1}
	
	Given a candidate sample $\bm\gamma^\dagger$, the mean $\mathbf x$ and variance $\bm\lambda$ of the density $f(\mathbf X|\mathbf{Y})$ are approximated via the following recursion:
	\begin{table}[!h]
		\begin{tabular}{lll}
			$(n = 0)$ &  & $\hat{\mathbf{x}}_{I\times 1}^\timeN{n} = \mathbf{0}_{I\times 1}$ \\[0.25cm]
			& & $\hat{\bm\lambda}_{I\times 1}^\timeN{n} = \mathbf{1}_{I\times 1}$ \\[0.8cm]
			$(n > 0)$ &  & $\overline{\mathbf{x}}_{I\times 1}^\timeN{n} = \hat{\mathbf{x}}_{I\times 1}^\timeN{n-1}$ \\[0.25cm]
			& & $\overline{\bm\lambda}_{I\times 1}^\timeN{n} = \hat{\bm\lambda}_{I\times 1}^\timeN{n-1}$\\[0.5cm]
			& & $\hat{\mathbf{y}}_{JI\times 1} = \mathcal{G}\Big( \overline{\mathbf{x}}_{I\times 1}^\timeN{n}, \mathbf{Z}_{J\times K} \cdot \bm\gamma^\dagger_{K\times 1} \Big)$\\[0.25cm]
			& & $\bm\sigma_{JI\times 1} = \Big(\mathbf{I}_{J\times J} \otimes \overline{\bm\lambda}_{I\times 1}^\timeN{n}\Big)\mathbf{1}_{J\times 1} + 1 \oslash \sqrt{\exp^\dagger\Big(\mathbf d_{JI\times 1}^\timeN{n}}\Big)$\\[0.25cm]
			& & $\mathbf{K}_{JI\times 1} = \Big(\mathbf{I}_{J\times J} \otimes \overline{\bm\lambda}_{I\times 1}^\timeN{n}\Big)\mathbf{1}_{J\times 1} \oslash \bm\sigma_{JI\times 1}$\\[0.75cm]
			& & $\hat{\mathbf{x}}_{I\times 1}^\timeN{n} =  \overline{\mathbf{x}}_{I\times 1}^\timeN{n} + \Bigg(\Big((\mathbf{y}_{JI\times 1} - \hat{\mathbf{y}}_{JI\times 1}) \odot \mathbf{K}_{JI\times 1}\Big)^T \mathbf A_{JI\times I}\Bigg)^T$\\[0.25cm]		 
			& & $\hat{\bm\lambda}_{I\times 1}^\timeN{n} = \overline{\bm\lambda}_{I\times 1}^\timeN{n} + \Bigg(\Big( \mathbf{K}_{JI\times 1} \odot \bm\sigma_{JI\times 1} \odot \mathbf{K}_{JI\times 1} \Big)^T\mathbf A_{JI\times I}\Bigg)^T$
		\end{tabular}
	\end{table}
	
	where $\otimes$ is the Kronecker product, $\odot$ the (element-wise) Hadamard product, $\oslash$ the (element-wise) Hadamard division, whereas $\mathbf{A} = \mathbf{I}_{I\times I} \otimes \big( n \mathbf{1}_{J\times 1} \big)$ is a \textit{scaling matrix} with $n=1/J$. As a byproduct of the Kalman filter, the marginal likelihood $f(\bm\gamma^\dagger|\mathbf Y)$ is multivariate Gaussian with mean $\hat{\mathbf{y}}$ and variance $\text{diag}(\bm\sigma)$, with $\text{diag}()$ being the linear operator that transforms a vector into a diagonal matrix. Finally, the array $\hat{\mathbf{X}}_{I\times N}$ contains the filtered latent states implied by the model whereas $\hat{\boldsymbol{\Lambda}}_{I\times N}$ is the array of variances associated with the filtered states. The smoothing part of the algorithm is implemented using the fixed-interval Kalman smoother \cite{sarkka:2013} where the filtered arrays $\hat{\mathbf{X}}_{I\times N}$ and $\hat{\bm\Lambda}_{I\times N}$ are used as input of the backward recursion.
	
\end{appendix}

\clearpage
\bibliographystyle{plain}
\bibliography{biblio}

\begin{thebibliography}{10}

\bibitem{andrieu+al:2010}
Christophe Andrieu, Arnaud Doucet, and Roman Holenstein.
\newblock Particle markov chain monte carlo methods.
\newblock {\em Journal of the Royal Statistical Society: Series B (Statistical
  Methodology)}, 72(3):269--342, 2010.

\bibitem{barca+pezzullo:2012}
Laura Barca and Giovanni Pezzulo.
\newblock Unfolding visual lexical decision in time.
\newblock {\em PloS one}, 7(4):e35932, 2012.

\bibitem{brockwell+al:2004}
Anthony~E Brockwell, Alex~L Rojas, and RE~Kass.
\newblock Recursive bayesian decoding of motor cortical signals by particle
  filtering.
\newblock {\em Journal of Neurophysiology}, 91(4):1899--1907, 2004.

\bibitem{calcagni+al:2017}
Antonio Calcagn{\`\i}, Luigi Lombardi, and Simone Sulpizio.
\newblock Analyzing spatial data from mouse tracker methodology: An entropic
  approach.
\newblock {\em Behavior research methods}, 49(6):2012--2030, 2017.

\bibitem{calenge2006package}
Cl{\'e}ment Calenge.
\newblock The package adehabitat for the r software: a tool for the analysis of
  space and habitat use by animals.
\newblock {\em Ecological modelling}, 197(3-4):516--519, 2006.

\bibitem{carpenter+al:2017}
Bob Carpenter, Andrew Gelman, Matthew~D Hoffman, Daniel Lee, Ben Goodrich,
  Michael Betancourt, Marcus Brubaker, Jiqiang Guo, Peter Li, and Allen
  Riddell.
\newblock Stan: A probabilistic programming language.
\newblock {\em Journal of statistical software}, 76(1), 2017.

\bibitem{Coco+al:2015}
Moreno Coco and Nicholas Duran.
\newblock {\em \pkg{mousetrap}: Process and Analyze Mouse-Tracking Data}, 2015.
\newblock {R} package version 1.0.0.

\bibitem{coco+duran:2016}
Moreno~I Coco and Nicholas~D Duran.
\newblock When expectancies collide: Action dynamics reveal the interaction
  between stimulus plausibility and congruency.
\newblock {\em Psychonomic bulletin \& review}, 23(6):1920--1931, 2016.

\bibitem{freeman:2017}
Jonathan~B Freeman.
\newblock Doing psychological science by hand.
\newblock {\em Current directions in psychological science}, In press:1--27,
  2017.

\bibitem{freeman+al:2010}
Jonathan~B Freeman and Nalini Ambady.
\newblock Mousetracker: Software for studying real-time mental processing using
  a computer mouse-tracking method.
\newblock {\em Behavior Research Methods}, 42(1):226--241, 2010.

\bibitem{frick+al:2017}
H~Frick and I~Kosmidis.
\newblock {\em \pkg{trackeR}: Infrastructure for Running and Cycling Data from
  GPS-Enabled Tracking Devices}, 2017.
\newblock {R} package version 1.0.0.

\bibitem{bayesplot:2018}
J~Gabry and T~Mahr.
\newblock {\em \pkg{bayesplot}: Plotting for Bayesian Models}, 2018.
\newblock {R} package version 1.6.0.

\bibitem{gelman+al:2014}
Andrew Gelman, John~B Carlin, Hal~S Stern, David~B Dunson, Aki Vehtari, and
  Donald~B Rubin.
\newblock {\em Bayesian data analysis}, volume~2.
\newblock CRC press Boca Raton, FL, 2014.

\bibitem{giorgino2009computing}
Toni Giorgino et~al.
\newblock Computing and visualizing dynamic time warping alignments in r: the
  dtw package.
\newblock {\em Journal of statistical Software}, 31(7):1--24, 2009.

\bibitem{hehman+al:2014}
Eric Hehman, Ryan~M Stolier, and Jonathan~B Freeman.
\newblock Advanced mouse-tracking analytic techniques for enhancing
  psychological science.
\newblock {\em Group Processes \& Intergroup Relations}, 18(3):384--401, 2015.

\bibitem{helske+vihola:2018}
J~Helske and M~Vihola.
\newblock {\em \pkg{bssm}: Bayesian Inference of Non-Linear and Non-Gaussian
  State Space Models}, 2018.
\newblock {R} package version 0.1.6-1.

\bibitem{helske:2017}
Jouni Helske.
\newblock Kfas: Exponential family state space models in {R}.
\newblock {\em Journal of Statistical Software}, 78(10), 2017.

\bibitem{kiers1997techniques}
Henk~AL Kiers.
\newblock Techniques for rotating two or more loading matrices to optimal
  agreement and simple structure: A comparison and some technical details.
\newblock {\em Psychometrika}, 62(4):545--568, 1997.

\bibitem{Kieslich+al:2018}
Pascal Kieslich, Dirk~U. Wulff, Felix Henninger, and Jonas M.~B. Haslbeck.
\newblock {\em \pkg{mousetrap}: Process and Analyze Mouse-Tracking Data}, 2018.
\newblock {R} package version 3.1.1.

\bibitem{kieslich+henninger:2017}
Pascal~J Kieslich and Felix Henninger.
\newblock Mousetrap: An integrated, open-source mouse-tracking package.
\newblock {\em Behavior research methods}, 49(5):1652--1667, 2017.

\bibitem{kranstauber+al:2017}
B~Kranstauber, M~Smolla, and AK~Scharf.
\newblock {\em \pkg{move}: Visualizing and Analyzing Animal Track Data}, 2017.
\newblock {R} package version 3.0.1.

\bibitem{kruschke:2014}
John Kruschke.
\newblock {\em Doing Bayesian data analysis: A tutorial with R, JAGS, and
  Stan}.
\newblock Academic Press, 2014.

\bibitem{mcneish+dumas:2017}
Daniel McNeish and Denis Dumas.
\newblock Nonlinear growth models as measurement models: A second-order growth
  curve model for measuring potential.
\newblock {\em Multivariate behavioral research}, 52(1):61--85, 2017.

\bibitem{schulte-mecklenbeck+al:2019}
Anton~Kuehberger Michael Schulte-Mecklenbeck and Joseph~G. Johnson.
\newblock {\em A Handbook of Process Tracing Methods}.
\newblock Routledge, New York, 2nd edition, 2019.

\bibitem{monaro+al:2017}
Merylin Monaro, Luciano Gamberini, and Giuseppe Sartori.
\newblock The detection of faked identity using unexpected questions and mouse
  dynamics.
\newblock {\em PloS one}, 12(5):e0177851, 2017.

\bibitem{pastore:2018}
Massimiliano Pastore.
\newblock Overlapping: a {R} package for estimating overlapping in empirical
  distributions.
\newblock {\em The Journal of Open Source Software}, 3(32):1023, 2018.

\bibitem{pebesma+al:2015}
E~Pebesma and B~Klus.
\newblock {\em \pkg{trajectories}: Classes and Methods for Trajectory Data},
  2015.
\newblock {R} package version 0.1-4.

\bibitem{ruitenberg+al:2016}
Marit~FL Ruitenberg, Wout Duthoo, Patrick Santens, Rachael~D Seidler, Wim
  Notebaert, and Elger~L Abrahamse.
\newblock Sequence learning in parkinson's disease: Focusing on action dynamics
  and the role of dopaminergic medication.
\newblock {\em Neuropsychologia}, 93:30--39, 2016.

\bibitem{sarkka:2013}
Simo S{\"a}rkk{\"a}.
\newblock {\em Bayesian filtering and smoothing}, volume~3.
\newblock Cambridge University Press, 2013.

\bibitem{shumway+stoffer:2006}
Robert~H Shumway and David~S Stoffer.
\newblock {\em Time series analysis and its applications: with R examples}.
\newblock Springer Science \& Business Media, 2006.

\bibitem{rstan:2018}
Development~Team Stan.
\newblock {\em \pkg{rstan}: the {R} interface to {Stan}}, 2018.
\newblock {R} package version 2.18.2.

\bibitem{shinystan:2018}
Development~Team Stan.
\newblock {\em \pkg{shinystan}: Interactive Visual and Numerical Diagnostics
  and Posterior Analysis for Bayesian Models}, 2018.
\newblock {R} package version 2.5.0.

\bibitem{stolier+freeman:2017}
Ryan~M Stolier and Jonathan~B Freeman.
\newblock A neural mechanism of social categorization.
\newblock {\em Journal of Neuroscience}, 37(23):5711--5721, 2017.

\bibitem{ggmcmc:2016}
Fernandez i~Marin Xavier.
\newblock {\em \pkg{ggmcmc}: Tools for Analyzing MCMC Simulations from Bayesian
  Inference}, 2016.
\newblock {R} package version 1.1.

\end{thebibliography}

\end{document}